\documentclass[conference]{IEEEtran}
\IEEEoverridecommandlockouts
\usepackage{cite}
\usepackage{amsmath,amssymb,amsfonts}
\usepackage{graphicx}
\usepackage{xcolor}
\usepackage{tikz}
\usetikzlibrary{arrows.meta,positioning,calc,fit,shapes.geometric}

\usepackage{multirow}
\usepackage{booktabs}
\usepackage{siunitx}
\usepackage{tabularx}
\usepackage{comment}
\usepackage{pifont}

\usepackage{algorithmic}
\usepackage{textcomp}
\usepackage{url}
\usepackage{balance}
\usepackage{hyperref}

\usepackage{xcolor}

\def\BibTeX{{\rm B\kern-.05em{\sc i\kern-.025em b}\kern-.08em
    T\kern-.1667em\lower.7ex\hbox{E}\kern-.125emX}}
\begin{document}


\title{Fair-Gate: Fairness-Aware Interpretable Risk Gating for Sex-Fair Voice Biometrics\\}


\author{\IEEEauthorblockN{Yangyang Qu}
\IEEEauthorblockA{\textit{EURECOM} \\
Sophia Antipolis, France \\
quy@eurecom.fr}
\and
\IEEEauthorblockN{Massimiliano Todisco}
\IEEEauthorblockA{\textit{EURECOM} \\
Sophia Antipolis, France \\
todisco@eurecom.fr}
\and
\IEEEauthorblockN{Chiara Galdi}
\IEEEauthorblockA{\textit{EURECOM} \\
Sophia Antipolis, France \\
galdi@eurecom.fr}
\and
\IEEEauthorblockN{Nicholas Evans}
\IEEEauthorblockA{\textit{EURECOM} \\
Sophia Antipolis, France \\
evans@eurecom.fr}
}

\maketitle

\begingroup
\makeatletter
\renewcommand\@makefnmark{}
\renewcommand\@makefntext[1]{\noindent #1}
\makeatother

  
\endgroup




\begin{abstract}
Voice biometric systems can exhibit sex-related performance gaps even when overall verification accuracy is strong. 
We attribute these gaps to two practical mechanisms: (i) demographic shortcut learning, where speaker classification training exploits spurious correlations between sex and speaker identity, and (ii) feature entanglement, where sex-linked acoustic variation overlaps with identity cues and cannot be removed without degrading speaker discrimination. 
We propose \textbf{Fair-Gate}, a fairness-aware and interpretable risk-gating framework that addresses both mechanisms in a single pipeline. 
Fair-Gate applies risk extrapolation to reduce variation in speaker-classification risk across proxy sex groups, and introduces a local complementary gate that routes intermediate features into an identity branch and a sex branch. 
The gate provides interpretability by producing an explicit routing mask that can be inspected to understand which features are allocated to identity versus sex-related pathways. 
Experiments on VoxCeleb1 show that Fair-Gate improves the utility--fairness trade-off, yielding more sex-fair ASV performance under challenging evaluation conditions.
\end{abstract}


\section{Introduction}
\label{sec:intro}

Recent progress in deep speaker embedding architectures has led to substantial gains in automatic speaker verification (ASV), a core technology for voice biometrics. 
Despite these improvements, ASV systems can still exhibit systematic performance differences across demographic groups, especially sex. 
Although subgroup performance may appear similar when each group is evaluated at its own operating point, such reporting can mask disparities that emerge when deploying a single global decision threshold shared by all users, which is the common practical setting~\cite{toussaint2021svevafair,hutiri24_interspeech}. 
Recent studies have documented these disparities and examined how dataset composition, demographic imbalance, and evaluation design affect fairness, commonly defined as comparable verification error rates across groups under a shared operating point~\cite{fenu21_fairvoice,chen2022scirep,hajavi2023study,estevez23_icassp,toussaint2021svevafair,hutiri24_interspeech,chouchane2024comparison,hutiri2022design}. 
Importantly, these effects persist under standard protocols and operating points, suggesting systematic group-dependent model behavior rather than evaluation artifacts~\cite{toussaint2021svevafair,hutiri24_interspeech}.

A common mitigation strategy is to reduce the extent to which ASV embeddings encode sensitive attributes such as sex. 
This is often implemented through adversarial objectives or auxiliary prediction losses, sometimes combined with reweighting or group-aware fusion~\cite{ganin2016domain,bhattacharya2019adapting,peri2023biasmitigation,jin22_arw,shen22_gfn,chouchanebio24}. 
However, in ASV, sex-linked acoustic cues are not purely nuisance factors: pitch, timbre, and resonance correlate with sex but also carry identity-relevant information. 
Consequently, enforcing strong sex invariance can suppress useful speaker cues and degrade verification performance~\cite{peri2023biasmitigation}. 
Related trade-offs have also been observed in privacy-oriented speaker anonymization and sex obfuscation, where reducing sex information can harm downstream utility~\cite{tomashenko2020introducing,tomashenko2022voiceprivacy,qu2025_raso}. 
This motivates a different goal: rather than enforcing global sex invariance, we seek to control where sex-linked variation is represented and how strongly it perturbs verification behavior under a shared decision threshold.

\begin{figure}[t]
    \centering
    \includegraphics[width=\linewidth]{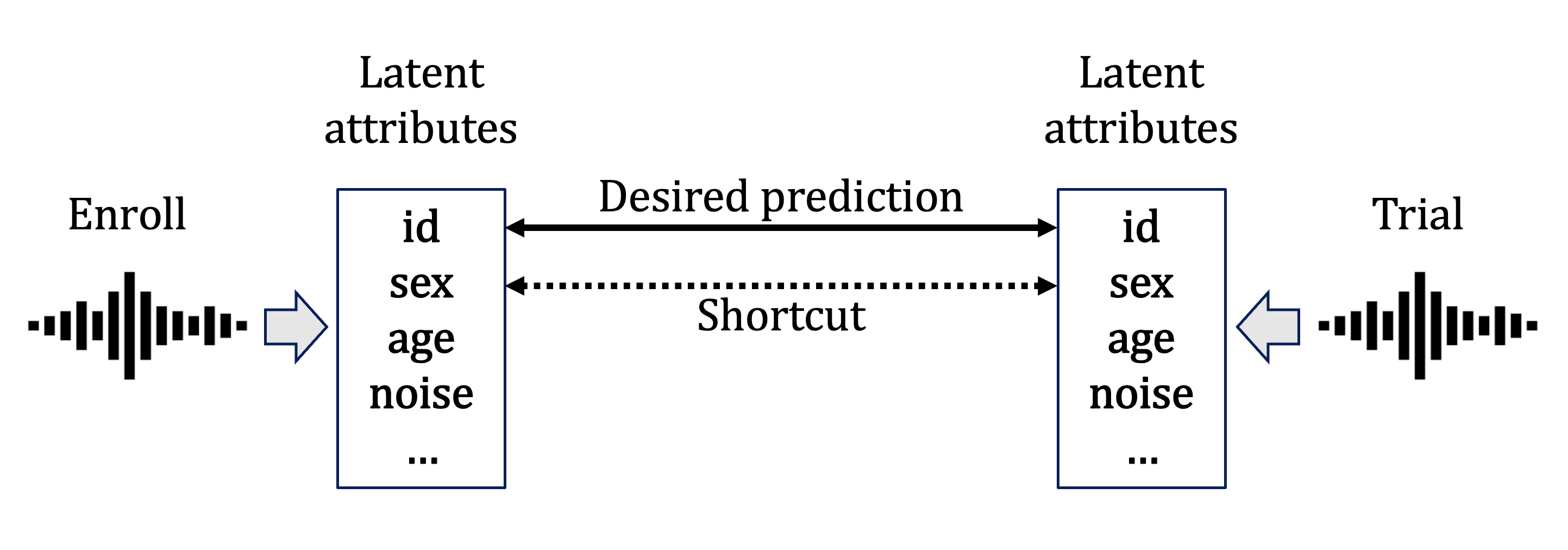} %
    \caption{Desired decision vs. demographic shortcut in speaker verification under a shared threshold. 
    A verifier should base its decision on identity evidence by comparing enrollment and test utterances (solid arrow). However, because sex affects acoustics (e.g., $F_0$ and formant structure) and can be spuriously correlated with speaker identity in the training data, the model may also exploit sex-linked cues as a shortcut (dashed arrow). Such shortcut reliance can shift score distributions differently for male and female speakers, leading to subgroup error-rate gaps when deploying a single global decision threshold.}

    \label{fig:causal_bias}
\end{figure}

Inspired by causal fairness formulations~\cite{kusner2017counterfactual}, we distinguish (i) the causal effect of sex on speech acoustics (e.g., $F_0$ and formants) from (ii) dataset-induced correlations between sex and speaker identity in the training data.  
As illustrated in Fig.~\ref{fig:causal_bias}, standard discriminative training can exploit such correlations as demographic shortcuts, shifting score distributions differently across groups, and producing unequal subgroup error rates under a shared threshold. 

\begin{figure*}[t]
    \centering
\includegraphics[width=0.83\textwidth]{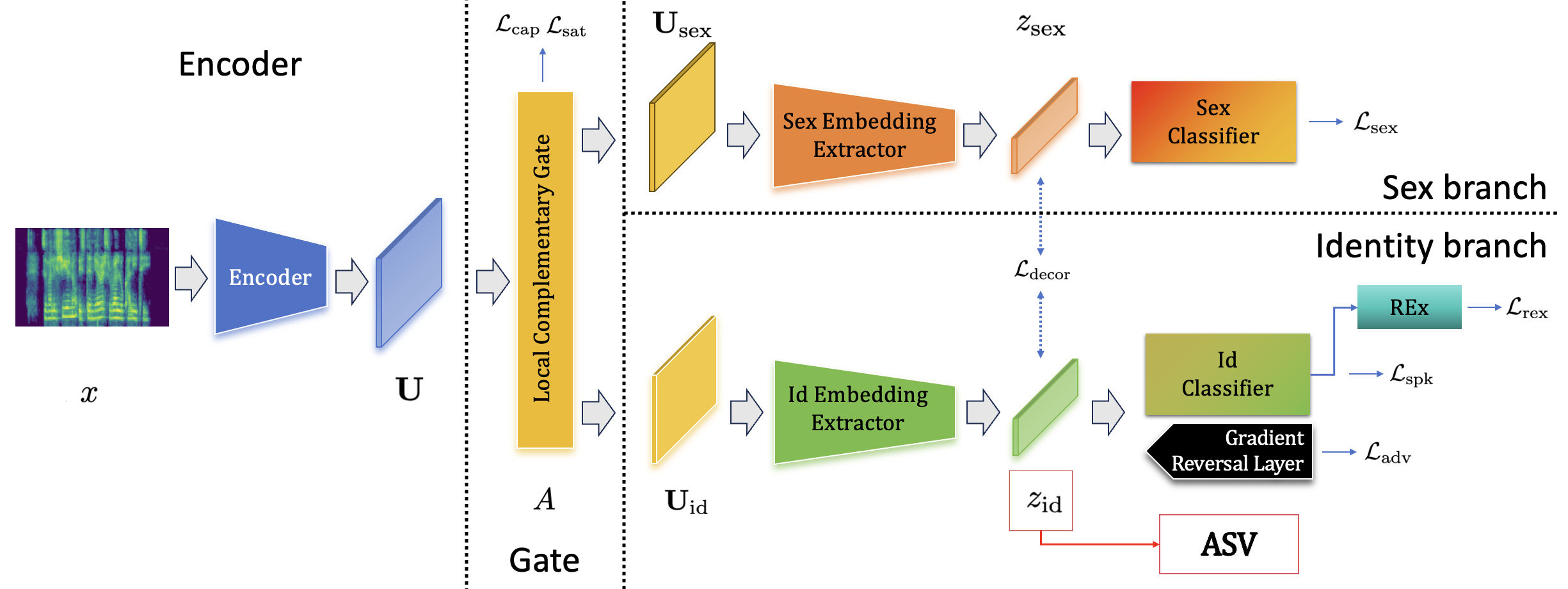}
    \caption{Overview of Fair-Gate. 
    The encoder produces frame-level features $\mathbf{U}$, which are complementarily soft-routed by a local mask $A$ (gate) into an identity branch and a sex branch. The identity branch produces the embedding $z_{\mathrm{id}}$, which is the only representation used for automatic speaker verification (ASV) at inference. 
    During training, the sex branch learns a sex embedding $z_{\mathrm{sex}}$ and predicts proxy sex labels $\hat{s}$ via a sex classifier ($\mathcal{L}_{\mathrm{sex}}$). The identity branch is optimized for speaker classification ($\mathcal{L}_{\mathrm{spk}}$), regularized by Risk Extrapolation (REx) across proxy sex groups ($\mathcal{L}_{\mathrm{rex}}$), and constrained by an adversarial sex classifier implemented through a Gradient Reversal Layer (GRL) ($\mathcal{L}_{\mathrm{adv}}$). A decorrelation loss ($\mathcal{L}_{\mathrm{decor}}$) encourages separation between $z_{\mathrm{id}}$ and $z_{\mathrm{sex}}$, while gate regularizers ($\mathcal{L}_{\mathrm{cap}}$, $\mathcal{L}_{\mathrm{sat}}$) prevent degenerate routing.}
    \label{fig:model}
\end{figure*}

This analysis highlights two fairness-relevant failure modes: 
(i) \emph{shortcut learning}, where speaker classification relies on sex-conditioned correlations that do not generalize equally across groups;  
(ii) \emph{feature entanglement}, where sex-related variation mixes with identity cues in the embedding used at inference, making it difficult to reduce subgroup error gaps without sacrificing verification performance.

Motivated by these mechanisms, we propose \textbf{Fair-Gate}, a unified training framework that improves the utility--fairness trade-off by addressing shortcut learning and feature entanglement in a single pipeline. In the following, we use the term \textit{utility} to refer specifically to verification performance, in order to clearly distinguish it from fairness performance.
First, we apply Risk Extrapolation (REx)~\cite{krueger2021out} across proxy sex groups to penalize differences in speaker-classification risk between groups, discouraging group-specific shortcuts. 
Second, we introduce a complementary local soft-routing gate that partitions intermediate features into two additive components routed to an identity branch and a sex branch, while preserving the original feature dimensionality. 
The sex branch provides an explicit pathway for sex-linked variation during training, reducing its leakage into the identity embedding used for verification at inference.\\

\noindent Our contributions are:
\begin{itemize}
\item We provide a causal analysis of sex-related bias in ASV, separating inherent sex-linked acoustic variation from dataset-induced correlations.
\item We propose Fair-Gate, combining Risk Extrapolation (REx) across proxy sex groups with a complementary local gating mechanism and branch-specific objectives to limit sex leakage into the deployed embedding.
\item We demonstrate improved utility--fairness trade-offs on VoxCeleb, and provide ablations that clarify the roles of complementary routing and branch-specific training objectives.
\end{itemize}

Despite sex and gender terms being commonly used interchangeably, there is a generally, though not universally accepted distinction. We adopt the terminology in~\cite{gauthier} and expressly avoid references to gender which refers to socially constructed roles and behaviour. Instead, we refer only to sex, which refers to biological attributes~\cite{prince}.
\section{Fair-Gate Framework}
\label{sec:method}

Fair-Gate extends a standard ECAPA-style speaker verification pipeline with complementary feature routing and fairness-aware training objectives. 
The framework illustrated in Fig.~\ref{fig:model} consists of three key components: 
(i) a shared encoder that extracts frame-level representations, 
(ii) a local complementary gate that routes these representations into an identity branch and a sex branch, and (iii) branch-specific objectives, including risk variance equalization, that promote speaker discrimination while reducing sex-dependent disparities under a shared operating threshold.\\

At inference, only the identity branch is retained for verification. 
Given an enrollment--test pair $(x_a,x_b)$, the verification score is computed using cosine similarity between identity embeddings:
\begin{equation}
\mathrm{score}(x_a,x_b)
= \cos\!\big(z_{\mathrm{id}}(x_a), z_{\mathrm{id}}(x_b)\big).
\end{equation}
A single shared threshold is then applied across all users.

\subsection{Encoder}
\label{sec:method_encoder}

Given an input log-Mel spectrogram $x \in \mathbb{R}^{F \times T}$, where $F$ is the number of Mel-frequency bins and $T$ the number of frames, the encoder $E(\cdot)$ produces frame-level features in $\mathbf{U} \in \mathbb{R}^{C \times T}$, where $C$ denotes the number of feature channels.

\subsection{Local Complementary Gating}
\label{sec:method_gate}
The local complementary gate (center of Fig.~\ref{fig:model}) softly allocates intermediate features between an identity branch and a sex branch while preserving dimensionality. 
Unlike global channel attention mechanisms, the gate operates at each time--channel location, enabling fine-grained routing. 
Preserving dimensionality is important here because the goal is information reallocation rather than compression: the gate should redistribute partially entangled cues between branches without imposing a fixed feature split or discarding speaker-discriminative content a priori.

\subsubsection{Mask computation and routing}

Given frame-level features $\mathbf{U}$, we compute a soft mask
\begin{equation}
A = \sigma\!\big(\mathrm{DWConv}_t(\mathbf{U})\big),
\end{equation}
where $\mathrm{DWConv}_t$ is a depthwise temporal convolution applied independently per channel and configured to preserve temporal length, and $\sigma(\cdot)$ is the sigmoid. 
Features are routed complementarily as
\begin{equation}
\label{eq:routing}
\mathbf{U}_{\mathrm{id}} = A \odot \mathbf{U}, 
\qquad
\mathbf{U}_{\mathrm{sex}} = (1-A) \odot \mathbf{U}.
\end{equation}

Both branches retain dimensionality and satisfy 
$\mathbf{U}_{\mathrm{id}} + \mathbf{U}_{\mathrm{sex}} = \mathbf{U}$ element-wise. 
This additive decomposition ensures that the gate reallocates information between branches, allowing Fair-Gate to learn where information should be represented rather than forcing identity- and sex-related cues into fixed disjoint subspaces.

\subsubsection{Gate regularization}
To avoid degenerate routing (e.g., collapsing all features to one branch or producing ambiguous allocations), we regularize the mask $A$ through two complementary terms defined below: a routing-mass control term $\mathcal{L}_{\mathrm{cap}}$ and a saturation term $\mathcal{L}_{\mathrm{sat}}$.
\subsubsection{Routing mass control}
Let
\[
\bar{a} = \frac{1}{BCT}\sum_{i,c,t} A_{i,c,t}.
\]
We regulate the average routing mass via
\[
\mathcal{L}_{\mathrm{cap}} = (\bar{a}-\rho_{\mathrm{id}})^2,
\]
where $\rho_{\mathrm{id}}$ controls the desired proportion of features allocated to the identity branch.

\subsubsection{Saturation constraint}
To encourage confident (near-binary) routing,
\[
\mathcal{L}_{\mathrm{sat}}
= \frac{1}{BCT}\sum_{i,c,t} A_{i,c,t}(1-A_{i,c,t}).
\]

\subsection{Identity and Sex Branches Objectives}
\label{sec:method_branches}

Each routed feature sequence is mapped to an utterance-level embedding using a branch-specific embedding extractor (Fig.~\ref{fig:model}):
\[
z_{\mathrm{id}} = f_{\mathrm{spk}}(\mathbf{U}_{\mathrm{id}}),
\qquad
z_{\mathrm{sex}} = f_{\mathrm{sex}}(\mathbf{U}_{\mathrm{sex}}).
\]
Here, $f_{\mathrm{spk}}(\cdot)$ and $f_{\mathrm{sex}}(\cdot)$ denote the identity- and sex-branch utterance-level embedding extractors, respectively. 
The identity extractor uses attentive statistics pooling~\cite{okabe2018attentive} to form $z_{\mathrm{id}}$, which is the only embedding used for ASV at inference. 
The sex branch is used only during training to capture sex-related variation and reduce its leakage into $z_{\mathrm{id}}$.

We do not use human-annotated sex labels. 
Instead, we obtain binary proxy sex labels $\hat{s}\in\{\mathrm{M},\mathrm{F}\}$ from a frozen pre-trained classifier and use them only during training. 
These proxy labels supervise the sex-classification objective, the adversarial sex objective, and the proxy-group partition used by REx. 
At inference, neither the proxy labels nor the sex branch is required.

Fair-Gate jointly optimizes speaker discrimination and fairness through branch-specific objectives. 
We denote cross-entropy by $\mathrm{CE}(\cdot,\cdot)$.

\subsubsection{Speaker classification}

The identity embedding is trained with an AAM-Softmax classifier:
\begin{equation}
\label{eq:Lspk}
\mathcal{L}_{\mathrm{spk}}
= \mathbb{E}_{(x,y)}
\mathrm{CE}\big(h_{\mathrm{spk}}(z_{\mathrm{id}}(x)),y\big),
\end{equation}
where $y$ denotes the speaker label and $h_{\mathrm{spk}}(\cdot)$ the speaker-classification head. 

\subsubsection{Adversarial constraint}

To discourage encoding of sex information in $z_{\mathrm{id}}$, 
we attach an adversarial sex classifier via a Gradient Reversal Layer (GRL)~\cite{ganin2016domain}:
\begin{equation}
\label{eq:Ladv}
\mathcal{L}_{\mathrm{adv}}
= \mathbb{E}_{(x,\hat{s})}
\mathrm{CE}\!\big(h^{\mathrm{adv}}_{\mathrm{sex}}(\mathrm{GRL}_\gamma(z_{\mathrm{id}}(x))),\hat{s}\big).
\end{equation}
Here, $h^{\mathrm{adv}}_{\mathrm{sex}}(\cdot)$ denotes the adversarial sex-classification head and $\mathrm{GRL}_\gamma(\cdot)$ a gradient reversal layer with reversal strength $\gamma$. 
Together, $\mathcal{L}_{\mathrm{spk}}$ and $\mathcal{L}_{\mathrm{adv}}$ encourage identity discrimination while reducing direct sex predictability in the deployed embedding.

\subsubsection{Sex classification}

The sex branch is trained to explicitly capture sex-related variation:
\begin{equation}
\label{eq:Lsex}
\mathcal{L}_{\mathrm{sex}}
= \mathbb{E}_{(x,\hat{s})}
\mathrm{CE}\big(h_{\mathrm{sex}}(z_{\mathrm{sex}}(x)),\hat{s}\big),
\end{equation}
where $h_{\mathrm{sex}}(\cdot)$ denotes the sex-classification head attached to $z_{\mathrm{sex}}$.  

\subsubsection{Embedding decorrelation}

To further reduce information overlap between branches, we penalize similarity between normalized embeddings:
\begin{equation}
\label{eq:Ldecor}
\mathcal{L}_{\mathrm{decor}}
= \mathbb{E}_{x}
\big\langle
\bar{z}_{\mathrm{id}}(x),
\bar{z}_{\mathrm{sex}}(x)
\big\rangle^2,
\quad
\bar{z}=\frac{z}{\|z\|_2}.
\end{equation}
This term does not enforce full statistical independence; rather, it discourages direct overlap between the two embeddings and complements the routing and adversarial objectives.

\subsubsection{Risk Variance Equalization}
\label{sec:method_rex}

Risk Extrapolation (REx) reduces discrepancies in speaker-classification risk loss across proxy sex groups.
Let $\mathcal{E}=\{\mathrm{M},\mathrm{F}\}$. 
The per-group risk is
\begin{equation}
\label{eq:RexRisk}
\mathcal{R}_e
= \mathbb{E}_{(x,y)\sim \mathcal{E}}
\mathrm{CE}\big(h_{\mathrm{spk}}(z_{\mathrm{id}}(x)),y\big),
\qquad
\bar{\mathcal{R}}
= \tfrac{1}{|\mathcal{E}|}\sum_{e\in\mathcal{E}}\mathcal{R}_e.
\end{equation}
The REx penalty is
\begin{equation}
\label{eq:Lrex}
\mathcal{L}_{\mathrm{rex}}
=\tfrac{1}{|\mathcal{E}|}\sum_{e\in\mathcal{E}}
(\mathcal{R}_e-\bar{\mathcal{R}})^2.
\end{equation}

Intuitively, if the speaker classifier relies on group-specific shortcuts, speaker-classification risk will differ systematically across proxy sex groups. 
Penalising the variance of $\{\mathcal{R}_e\}$, therefore, encourages the model to rely on speaker evidence that transfers more uniformly across groups.

In practice, group risks are estimated within mini-batches, and the penalty is applied only when both groups are sufficiently represented in the batch.
\subsection{Overall Training Objective}
\label{sec:method_total}

The full objective combines utility (i.e. verification performance), fairness, routing, and separation terms:
\begin{equation}
\label{eq:Ltotal-method}
\begin{split}
\mathcal{L}_{\mathrm{total}}
&= \mathcal{L}_{\mathrm{spk}}
+ \lambda_s \mathcal{L}_{\mathrm{sex}}
+ \lambda_{\mathrm{adv}} \mathcal{L}_{\mathrm{adv}}
+ \lambda_{\mathrm{decor}} \mathcal{L}_{\mathrm{decor}} \\
& \quad + \lambda_{\mathrm{cap}} \mathcal{L}_{\mathrm{cap}}
+ \lambda_{\mathrm{sat}} \mathcal{L}_{\mathrm{sat}}
+ \lambda_{\mathrm{rex}} \mathcal{L}_{\mathrm{rex}}.
\end{split}
\end{equation}

\begin{table}[t]
\centering
\small
\caption{VoxCeleb1 verification protocols used for evaluation.}
\label{tab:vox_protocols}
\begin{tabular}{lccp{3cm}}
\toprule
Protocol & Speakers & \# Trials & Description \\
\midrule
Vox1-O & 40   & 37,720  & Original test list \\
Vox1-E & 1,251 & 581,480 & Expanded trial list \\
Vox1-H & 1,251 & 552,536 & Hard (same nationality \& sex impostors) \\
\bottomrule
\end{tabular}
\end{table}

\begin{table*}[!t]
\centering
\small
\setlength{\tabcolsep}{3.0pt}
\renewcommand{\arraystretch}{0.95}
\caption{Results on VoxCeleb1 (O/E/H). We report EER [\%], minDCF, and sex fairness measured by GARBE at the threshold $\tau_{1\%}$.}

\label{tab:results_vox_best_ours_only_relayout}

\begin{tabular*}{\textwidth}{@{\extracolsep{\fill}} l ccc ccc ccc @{}}
\toprule
\multirow{2}{*}{Model}
& \multicolumn{3}{c}{\textbf{Vox1-O}}
& \multicolumn{3}{c}{\textbf{Vox1-E}}
& \multicolumn{3}{c}{\textbf{Vox1-H}}
\\
\cmidrule(lr){2-4}\cmidrule(lr){5-7}\cmidrule(lr){8-10}
& EER $\downarrow$ & minDCF $\downarrow$ & GARBE  $\downarrow$
& EER  $\downarrow$ & minDCF $\downarrow$ & GARBE  $\downarrow$
& EER  $\downarrow$ & minDCF $\downarrow$ & GARBE  $\downarrow$
\\
\midrule


ECAPA-TDNN~\cite{desplanques2020ecapa} &
1.12 & 0.14 & \textbf{0.16} &
1.34 & 0.17 & 0.11 &
2.62 & \textbf{0.25} & 0.10 \\
ECAPA-TDNN + GRL & 0.98 & 0.13 & 0.22 & 1.25 & 0.14 & 0.12 & 2.57 & 0.28 & 0.10 \\

VoxDisentangler~\cite{nam2024disentangled}
& \textbf{0.82} & 0.12 & 0.17
& 1.15 & 0.14 & 0.11
& 2.40 & 0.26 & 0.10
\\

\textbf{Fair-Gate (ours)} &
0.92 & \textbf{0.11} & 0.26 &
\textbf{1.11} & \textbf{0.14} & \textbf{0.05} &
\textbf{2.25} & 0.26 & \textbf{0.07} \\

\bottomrule
\end{tabular*}
\vspace{2pt}
\end{table*}

\begin{table*}[t]
\centering
\small
\setlength{\tabcolsep}{5.5pt}
\renewcommand{\arraystretch}{1.08}
\caption{Ablation results on Vox1-H. We report EER [\%], $\mathrm{GARBE}(\tau_{1\%})$, and subgroup-specific FNMR/FMR [\%] at the same shared threshold $\tau_{1\%}$. Bold indicates the best value in each column.}
\label{tab:ablation_vox1h_expanded}
\resizebox{\textwidth}{!}{%
\begin{tabular}{lccccc|cccccc}
\toprule
Setting & Cap & Sat & Gs & Adv & REx & EER$\downarrow$ & GARBE$\downarrow$ & FNMR$_M$ & FNMR$_F$ & FMR$_M$ & FMR$_F$ \\
\midrule
Main w/o Cap              & --         & \checkmark & \checkmark & \checkmark & \checkmark & 2.66 & 0.09 & 1.03 & \textbf{0.95} & 4.76 & 6.30 \\
Main w/o Sat              & \checkmark & --         & \checkmark & \checkmark & \checkmark & 2.30 & 0.07 & \textbf{0.95} & 1.07 & 3.98 & 4.68 \\
Main w/o Gs               & \checkmark & \checkmark & --         & \checkmark & \checkmark & 2.66 & 0.09 & 1.03 & \textbf{0.95} & 4.76 & 6.30 \\
Main w/o Adv              & \checkmark & \checkmark & \checkmark & --         & \checkmark & 2.27 & \textbf{0.07} & 0.96 & 1.06 & 3.81 & \textbf{4.48} \\
Main w/o REx              & \checkmark & \checkmark & \checkmark & \checkmark & --         & 2.55 & 0.08 & 0.96 & 1.06 & 4.54 & 5.60 \\
\textbf{Main, Full} & \checkmark & \checkmark & \checkmark & \checkmark & \checkmark & \textbf{2.25} & \textbf{0.07} & 0.96 & 1.07 & \textbf{3.80} & 4.49 \\
\bottomrule
\end{tabular}%
}
\end{table*}

\section{Experimental Setup and Metrics}

\subsection{Datasets and Protocols}

We train all models on the VoxCeleb2 development set~\cite{chung2018voxceleb2,nagrani2020voxceleb}, 
which contains over one million utterances from more than 6,000 speakers collected in unconstrained, real-world conditions. 
Evaluation is performed on VoxCeleb1~\cite{nagraniy2017voxceleb,nagrani2020voxceleb} 
using the official verification protocols Vox1-O, Vox1-E, and Vox1-H, ensuring comparability with prior ASV and fairness studies.

Vox1-O is the \emph{O}riginal test list, Vox1-E is an \emph{E}xpanded trial set, and Vox1-H is a \emph{H}ard protocol in which non-mated trials are constructed from speakers matched in nationality and proxy sex. 
These protocols differ in the number and difficulty of verification trials, as summarized in Table~\ref{tab:vox_protocols}.

\subsection{Metrics and baselines}
The evaluation of the speaker verification system for both utility and fairness is reported in terms of classification error analysis. In particular we observe:
\begin{itemize}
    \item[-] False match rate (FMR): proportion of the completed biometric non-mated comparison trials that result in a false match; 
    \item[-] False non-match rate (FNMR): proportion of the completed biometric mated comparison trials that result in a false non-match;
\end{itemize}

\vspace{.25cm}

For the sake of comparison with the state of the art, system's utility is reported in terms of the equal error rate (EER, where FMR=FNMR) and minimum detection cost function (minDCF)~\cite{martin2010nist} as well:

\begin{equation}
\label{eq:mindcf}
\begin{aligned}
\mathrm{DCF}(\tau)
&= C_{\mathrm{FNMR}}\,P_{\mathrm{target}}\,\mathrm{FNMR}(\tau) \\
&\quad + C_{\mathrm{FMR}}\,(1-P_{\mathrm{target}})\,\mathrm{FMR}(\tau), \\
\mathrm{minDCF}
&= \min_{\tau}\, \mathrm{DCF}(\tau).
\end{aligned}
\end{equation}
where $\tau$ is the decision threshold and $\mathrm{FMR}(\tau)$ / $\mathrm{FNMR}(\tau)$ are computed from pooled trials (male and female speakers); $P_{\mathrm{tar}}$ is the prior probability of a genuine attempt (target); and $C_{\mathrm{FNMR}}$ and $C_{\mathrm{FMR}}$ are weighting parameters that adjust the relative importance of false non-matches and false matches.
We use the standard NIST SRE setting with $P_{\mathrm{tar}}=0.01$ and $C_{\mathrm{FNMR}}=C_{\mathrm{FMR}}=1$.
Unlike the EER, where equal error importance is assumed and no prior information is provided, the minDCF measures the minimum expected application cost by incorporating error costs and class priors, and selecting the optimal threshold for that specific operating scenario.

Following prior ASV fairness evaluations~\cite{chouchane2024comparison}, we report fairness using GARBE~\cite{howard2022evaluating} at a fixed operating point specified by a threshold~$\tau$.

GARBE is defined as follows:
\begin{equation}
\label{eq:garbe}
\mathrm{GARBE}(\tau)
= \alpha\, G_{\mathrm{FMR}_\tau}
+ (1-\alpha)\, G_{\mathrm{FNMR}_\tau}
\end{equation}
where $G_{\mathrm{FMR}_\tau}$ and $G_{\mathrm{FNMR}_\tau}$ are the Gini coefficients 
computed over subgroup FMRs and FNMRs at threshold $\tau$, respectively. 
We set $\alpha=0.5$ so that FMR and FNMR disparities are equally weighted.

While identical thresholds are used for male and female score distributions for all experiments, different thresholds $\tau$ are computed according to the system evaluated (baselines and proposed) and the metric. For EER assessment, $\tau$ corresponds to the decision threshold where the system achieves $FMR=FNMR$. For the computation of $minDCF$, $\tau$ is selected such that $DCF$ is minimised. Finally, for fairness, the threshold for GARBE is fixed at $\tau=1\%$, such that $FMR(\tau)= 10^{-2}$, which we indicate in the following with $\mathrm{GARBE}(\tau_{1\%})$.


We compare our approach against three representative baselines covering utility-oriented training, adversarial invariance learning, and disentanglement-based modeling:

\begin{enumerate}

\item \textbf{ECAPA-TDNN}~\cite{desplanques2020ecapa}.
We adopt ECAPA-TDNN as a strong utility-oriented backbone. ECAPA-TDNN (Emphasized Channel Attention, Propagation and Aggregation Time Delay Neural Network) enhances the conventional TDNN architecture through channel-dependent frame attention, Res2Net multi-scale feature extraction, and attentive statistical pooling. These components improve the modeling of speaker-discriminative information while maintaining robustness to channel and acoustic variability. In our comparison, this model serves as the primary utility-driven baseline, optimized solely for speaker verification without any explicit fairness or invariance constraint.

\item \textbf{ECAPA-TDNN + GRL}.
To provide an adversarial invariance baseline, we augment ECAPA-TDNN with a gradient reversal layer (GRL)~\cite{ganin2016domain} to the speaker embedding. A sex classifier is trained adversarially through the GRL to discourage the encoding of sex-related information in the embedding space. During training, the speaker classification objective and the adversarial sex classification objective are jointly optimized, with the GRL reversing the gradient from the adversary to promote invariance. This setup follows the standard domain-adversarial learning paradigm and represents a commonly used strategy for mitigating demographic leakage while preserving speaker verification utility.


\item \textbf{VoxDisentangler}~\cite{nam2024disentangled}. VoxDisentangler is a disentanglement-based speaker verification framework that explicitly separates speaker identity from environmental variability (e.g., channel and noise conditions) into distinct latent representations. 
Shortcut learning in ASV is not limited to demographic attributes: models can also exploit spurious correlations between speaker identity and nuisance factors such as channel, background noise, or recording conditions. 
Although VoxDisentangler is not designed to disentangle demographic attributes such as sex, we include it as a baseline to assess whether disentangling non-identity nuisance factors can indirectly mitigate sex-dependent performance disparities.
\end{enumerate}

All methods are trained on the VoxCeleb2 development set and evaluated on VoxCeleb1 using the original (O), extended (E), and hard (H) trial protocols. To ensure a fair comparison, we follow identical training data splits, preprocessing steps, and evaluation metrics across all systems. Unless otherwise stated, all reported experiments use a batch size of 512.

\section{Results}
\label{sec:results}

\subsection{Results on VoxCeleb1-O/E/H.}
Results are presented in Table~\ref{tab:results_vox_best_ours_only_relayout}. 
GARBE results show that Fair-Gate improves sex-group fairness for Vox1-E/H while maintaining competitive utility.
For VoxCeleb1-O, Fair-Gate yields a larger sex disparity 
compared to competing approaches.
For the smallest and least challenging O protocol,
subgroup disparity estimates are more sensitive to trial composition and the 
precise operating point. In contrast, for the extended and harder Vox1-E and Vox1-H protocols, Fair-Gate consistently reduces sex disparity while maintaining strong utility, suggesting that risk equalization and the complementary gating mechanisms are beneficial under more challenging evaluation conditions where sex-dependent shortcut cues are more likely to be exploited.

For Vox1-E, Fair-Gate achieves the best fairness score of $\mathrm{GARBE}(\tau_{1\%})=0.05$,
substantially lower than for ECAPA (0.11), GRL (0.12), and VoxDisentangler (0.11).
Improved fairness comes with improved utility: Fair-Gate reduces the EER to 1.11\% (vs.\ 1.34\% ECAPA, 1.25\% GRL, 1.15\% VoxDisentangler),
and delivers a minDCF of 0.14 (vs.\ 0.17 for ECAPA, and the same results of 0.14 for GRL/VoxDisentangler).

For Vox1-H, Fair-Gate achieves the best EER and GARBE$(\tau_{1\%})$ results, while the minDCF is comparable to that of the strongest baselines. 
Notably, the GRL baseline does not improve fairness over ECAPA: for Vox1-E, GARBE increases from 0.11 to 0.12 whereas, for Vox1-H, the result of 0.10 suggests that adversarial invariance alone is insufficient to equalise subgroup error rates for a common operating point.
Overall, Fair-Gate offers the best fairness for Vox1-E/H demonstrating a stronger utility--fairness trade-off than the baselines.

 \subsection{Ablation study on VoxCeleb1-H}
Ablation results on Vox1-H are shown in Table~\ref{tab:ablation_vox1h_expanded}. Overall, the selected Fair-Gate configuration with moderate REx ($r=0.005$) provides the most favorable utility--fairness trade-off. It achieves the best EER (2.25\%) while remaining highly competitive on $\mathrm{GARBE}(\tau_{1\%})$ and subgroup-specific error rates. The largest degradation is observed when removing either the routing-mass regularizer or the sex-branch supervision: both \textit{w/o Cap} and \textit{w/o Gs} worsen from 2.25\% / 0.07 to 2.66\% / 0.09 in EER / GARBE, and substantially increase subgroup FMRs, especially for the female subgroup. This suggests that the main subgroup disparity on Vox1-H is driven primarily by the false-match side, and that complementary routing control together with explicit sex-branch supervision is important for limiting subgroup-dependent shortcut reliance.

The remaining ablations indicate that the other components play more limited or supporting roles. Removing REx still degrades the trade-off, increasing EER from 2.25\% to 2.55\% and GARBE from 0.07 to 0.08, while also raising subgroup FMRs from 3.80\% / 4.49\% to 4.54\% / 5.60\%, which shows that risk equalization contributes meaningfully under the shared operating point. By contrast, removing the adversarial term changes fairness only marginally but slightly worsens utility (2.25\% $\rightarrow$ 2.27\%), suggesting that adversarial invariance is not the primary source of subgroup-gap reduction. 

\section{Conclusions}
We present Fair-Gate, a fairness-oriented training framework for speaker verification that reduces discrepancies in sex-dependent error rates at common operating points.
Fair-Gate combines risk variance equalization across proxy sex groups to discourage group-specific shortcuts during speaker classification with complementary local gating to explicitly route intermediate features into identity and sensitive pathways, restricting the leakage of sex-linked variation into speaker embeddings.
Results derived using the VoxCeleb1 database shows that Fair-Gate improves the utility--fairness trade-off, with the clearest gains being achieved for the most challenging protocols.
Future work should explore more reliable proxy-group construction, extend the framework to additional sensitive attributes, and evaluate robustness under cross-corpus shifts and broader deployment conditions.

\section*{Acknowledgement}
This work was supported by the French Agence Nationale de la Recherche (ANR) via the SpeechPrivacy (ANR-23-CE23-0022) and GOOD-BIAS (ANR-25-CE39-6459) projects.

\balance
\bibliographystyle{IEEEtran}
\bibliography{mybib}

@inproceedings{krueger2021out,
  title={Out-of-distribution generalization via risk extrapolation (rex)},
  author={Krueger, David and Caballero, Ethan and Jacobsen, Joern-Henrik and Zhang, Amy and Binas, Jonathan and Zhang, Dinghuai and Le Priol, Remi and Courville, Aaron},
  booktitle={International conference on machine learning},
  pages={5815--5826},
  year={2021},
  organization={PMLR}
}

@article{ganin2016domain,
  title={Domain-adversarial training of neural networks},
  author={Ganin, Yaroslav and Ustinova, Evgeniya and Ajakan, Hana and Germain, Pascal and Larochelle, Hugo and Laviolette, Fran{\c{c}}ois and March, Mario and Lempitsky, Victor},
  journal={Journal of machine learning research},
  volume={17},
  number={59},
  pages={1--35},
  year={2016}
}

@inproceedings{nagraniy2017voxceleb,
  title={VoxCeleb: A large-scale speaker identification dataset},
  author={Nagraniy, Arsha and Chungy, Joon Son and Zisserman, Andrew},
  booktitle={Proc. Interspeech},
  pages={2616--2620},
  year={2017}
}

@inproceedings{chung2018voxceleb2,
  title={VoxCeleb2: Deep speaker recognition},
  author={Chung, J and Nagrani, A and Zisserman, A},
  booktitle={Proc. Interspeech},
  year={2018},
}

@inproceedings{hajavi2023study,
  title={A study on bias and fairness in deep speaker recognition},
  author={Hajavi, Amirhossein and Etemad, Ali},
  booktitle={Proc. ICASSP},
  pages={1--5},
  year={2023},
}

@inproceedings{hutiri2022design,
  title={Design Guidelines for Inclusive Speaker Verification Evaluation Datasets},
  author={Hutiri, Wiebke and Gorce, Lauriane and Ding, Aaron Yi},
  booktitle={Proc. Interspeech},
  pages={1293--1297},
  year={2022}
}

@inproceedings{bhattacharya2019adapting,
  title={Adapting end-to-end neural speaker verification to new languages and recording conditions with adversarial training},
  author={Bhattacharya, Gautam and Alam, Jahangir and Kenny, Patrick},
  booktitle={Proc. ICASSP},
  pages={6041--6045},
  year={2019}
  }

@inproceedings{desplanques2020ecapa,
  title={ECAPA-TDNN: Emphasized Channel Attention, Propagation and Aggregation in TDNN based speaker verification},
  author={Desplanques, Brecht and Thienpondt, Jenthe and Demuynck, Kris},
  booktitle={Proc. Interspeech},
  pages={3830--3834},
  year={2020}
}

@conference{chouchanebio24,
  author = {Chouchane, Oubaida and  Panariello, Michele and  Galdi, Chiara and  Todisco, Massimiliano and  Evans, Nicholas},
  title = {Fairness and privacy in voice biometrics: A study of gender influences using wav2vec 2.0},
  booktitle = {Proc. BIOSIG},
  year = {2023}
}

@conference{chouchane2024comparison,
  author = {Chouchane, Oubaida and  Busch, Christoph and  Galdi, Chiara and  Evans, Nicholas and  Todisco, Massimiliano},
  title = {A comparison of differential performance metrics for the evaluation of automatic speaker verification fairness},
  booktitle = {Proc. ODYSSEY},
  year = {2024}
}

@article{nagrani2020voxceleb,
  title={Voxceleb: Large-scale speaker verification in the wild},
  author={Nagrani, Arsha and Chung, Joon Son and Xie, Weidi and Zisserman, Andrew},
  journal={Computer Speech \& Language},
  volume={60},
  pages={101027},
  year={2020},
  publisher={Elsevier}
}

@article{toussaint2021svevafair,
  title={SVEva Fair: A Framework for Evaluating Fairness in Speaker Verification},
  author={Toussaint, Wiebke and Ding, Aaron Yi},
  journal={arXiv preprint arXiv:2107.12049},
  year={2021}
}

@inproceedings{hutiri24_Interspeech,
  title={As Biased as You Measure: Methodological Pitfalls of Bias Evaluations in Speaker Verification Research},
  author={Hutiri, Wiebke and Patel, Tanvina and Ding, Aaron Yi and Scharenborg, Odette},
  booktitle={Proc. Interspeech},
  year={2024},
  doi={10.21437/Interspeech.2024-1158}
}

@inproceedings{fenu21_fairvoice,
  title={Fair Voice Biometrics: Impact of Demographic Imbalance on Group Fairness in Speaker Recognition},
  author={Fenu, Gianni and Marras, Mirko and Medda, Giacomo and Meloni, Giacomo},
  booktitle={Proc. Interspeech},
  year={2021},
  pages={1892--1896},
  doi={10.21437/Interspeech.2021-1857}
}

@article{peri2023biasmitigation,
  title={A study of bias mitigation strategies for speaker recognition},
  author={Peri, Raghuveer and Somandepalli, Krishna and Narayanan, Shrikanth},
  journal={Computer Speech \& Language},
  volume={79},
  pages={101481},
  year={2023},
  doi={10.1016/j.csl.2022.101481}
}

@inproceedings{jin22_arw,
  title={Adversarial Reweighting for Speaker Verification Fairness},
  author={Jin, Minho and Ju, Chelsea J.-T. and Chen, Zeya and Liu, Yi-Chieh and Droppo, Jasha and Stolcke, Andreas},
  booktitle={Proc. Interspeech},
  year={2022},
  pages={4800--4804}
}

@inproceedings{shen22_gfn,
  title={Improving Fairness in Speaker Verification via Group-Adapted Fusion Network},
  author={Shen, Hua and Yang, Yuguang and Sun, Guoli and Langman, Ryan and Han, Eunjung and Droppo, Jasha and Stolcke, Andreas},
  booktitle={Proc. ICASSP},
  year={2022}
}

@inproceedings{estevez23_icassp,
  title={Study on the Fairness of Speaker Verification Systems Across Accent and Gender Groups},
  author={Est{\'e}vez, Mariel and Ferrer, Luciana},
 booktitle={Proc. ICASSP},
  pages={1--5},
  year={2023}
}

@article{chen2022scirep,
  title={Exploring racial and gender disparities in voice biometrics},
  author={Chen, Xingyu and Li, Zhengxiong and Setlur, Srirangaraj and Xu, Wenyao},
  journal={Scientific Reports},
  volume={12},
  number={1},
  pages={3723},
  year={2022},
  doi={10.1038/s41598-022-06673-y}
}

@inproceedings{qu2025_raso,
  title={Reference-free Adversarial Sex Obfuscation in Speech},
  author={Qu, Yangyang and Panariello, Michele and Todisco, Massimiliano and Evans, Nicholas},
  booktitle={APSIPA 2025, 17th Asia Pacific Signal and Information Processing Association Annual Summit and Conference},
  year={2025}
}

@inproceedings{nam2024disentangled,
  title={Disentangled Representation Learning for Environment-agnostic Speaker Recognition},
  author={Nam, KiHyun and Heo, Hee-Soo and Jung, Jee-weon and Chung, Joonson},
  booktitle={Proc. Interspeech 2024},
  pages={2130--2134},
  year={2024}
}

@article{tomashenko2022voiceprivacy,
  title={The voiceprivacy 2020 challenge: Results and findings},
  author={Tomashenko, Natalia and Wang, Xin and Vincent, Emmanuel and Patino, Jose and Srivastava, Brij Mohan Lal and No{\'e}, Paul-Gauthier and Nautsch, Andreas and Evans, Nicholas and Yamagishi, Junichi and O’Brien, Benjamin and others},
  journal={Computer Speech \& Language},
  volume={74},
  pages={101362},
  year={2022},
  publisher={Elsevier}
}

@inproceedings{tomashenko2020introducing,
  title={Introducing the VoicePrivacy initiative},
  author={Tomashenko, Natalia and Srivastava, Brij Mohan Lal and Wang, Xin and Vincent, Emmanuel and Nautsch, Andreas and Yamagishi, Junichi and Evans, Nicholas and Patino, Jose and Bonastre, Jean-Fran{\c{c}}ois and No{\'e}, Paul-Gauthier and others},
  booktitle={Proc. Interspeech},
  year={2020}
}

@inproceedings{okabe2018attentive,
  title={Attentive Statistics Pooling for Deep Speaker Embedding},
  author={Okabe, Koji and Koshinaka, Takafumi and Shinoda, Koichi},
  booktitle={Proc. Interspeech},
  pages={2252--2256},
  year={2018}
}

@article{kusner2017counterfactual,
  title={Counterfactual fairness},
  author={Kusner, Matt J and Loftus, Joshua and Russell, Chris and Silva, Ricardo},
  journal={Advances in neural information processing systems},
  volume={30},
  year={2017}
}

@inproceedings{howard2022evaluating,
  title={Evaluating proposed fairness models for face recognition algorithms},
  author={Howard, John J and Laird, Eli J and Rubin, Rebecca E and Sirotin, Yevgeniy B and Tipton, Jerry L and Vemury, Arun R},
  booktitle={Proc. ICPR},
  pages={431--447},
  year={2022}
}

@inproceedings{martin2010nist,
  title={The NIST 2010 speaker recognition evaluation.},
  author={Martin, Alvin F and Greenberg, Craig S},
  booktitle={Interspeech},
  pages={2726},
  year={2010}
}

@conference{gauthier,
  author = {Noe, Paul-Gauthier and  Mohammadamini, Mohammad and  Matrouf, Driss and  Parcollet, Titouan and  Nautsch, Andreas and  Bonastre, Jean-François},
  title = {Adversarial disentanglement of speaker representation for attribute-driven privacy preservation},
  booktitle = {Proc. Interspeech},
  year = {2021}
}

@article{prince,
author = {Virginia Prince},
title = {Sex vs. Gender},
journal = {International Journal of Transgenderism},
volume = {8},
number = {4},
pages = {29--32},
year = {2005},
publisher = {Taylor \& Francis}
}

\end{document}